\documentclass[UTF-8]{article}
\usepackage{graphicx}
\title{A novel type of Automata for dynamic, heterogeneous and random architectures}
\author{{\small Weijun ZHU}\\
{\footnotesize School of Information Engineering, Zhengzhou University, Zhengzhou, 450001, China}\\
{\footnotesize E-mail: zhuweijun@zzu.edu.cn}}

\begin{document}

\maketitle

\begin{abstract}
In this paper, the author aims to establish a mathematical model for a mimic computer. To this end, a novel automaton is proposed. First, a one-dimensional cellular automaton is used for expressing some dynamic changes in the structure of a computing unit, a sequential automaton is employed to describe some state transitions, a hierarchical automaton is employed to express the different granularities of some computing units, and a probabilistic automaton is used to depict some random changes of a computing unit. Second, the new automaton is obtained by combining the various types of automata mentioned above in the certain logical relationship. To the best of our knowledge, the new automaton model is the first automaton which can portray the operation semantics for a mimic computing system, and it can directly describe some behaviors of a mimic computer.
\end{abstract}
\section{Introduction}
It is well known that the Turing machine and many automata establish mathematical models for all kinds of computer systems in traditional computing. The von Neumann architecture solves the engineering and practical problems of the above computational theory [1]. However, with the continuous development of high-performance computers, power consumption is becoming one of the biggest stumbling block. In order to solve this problem, the high-efficiency computing technique has become a popular way of development of high-performance computers [1].

China developed the world's first prototype of mimic computers under the leadership of Academician Wu Jiangxing in September 2013 [2]. "Mimic computing aim to obtain the essence of the high-efficiency computing through the multidimensional-reconstruction-based functional (mimic variety) architecture." [1] "It aims to improve efficiency of running by applying the idea that the application determines the structure and the structure determines the performance, as well as by timely dynamic reconstructing the corresponding systemic structure or environment in execution." The test results from a third-party show that the ratio of the efficiency of mimic computing raise 13.6-315 times on more than 500 scenarios of web service, N-body and image recognition [2].

As far as automaton theory is concerned, it has been proposed for some decades. The early studies on automata provide a mathematical model which portrays some systemic behaviors, for the different computing systems which have the different expressive abilities. What is more, these studies also provides a theoretical basis and a formal tool for the development of compilers. In recent years, various automata have been applied to model checking and formal verification. And these automata have been widely used in a series of fields such as CPU design, network protocol verification, security protocol verification and software engineering analysis.

As for Mimic Computing (MC), none of the existing automata can establish a formal model of mimic computers. We need a model and method of automata. Motivated by it, a new automaton called Mimic Automaton (MA) is presented in this paper. On this basis, a formal model of mimic computing is established. Furthermore, a Mimic Model Checking (MMC) method is proposed for Mimic Formal Verifications (MFV). In addition, the MMC can be expected to detect some malwares on mimic computers. These are the contributions of this paper.

\section{The definition of the mimic automata}
The definition of a mimic automaton is based on the Sequential Automata (SA) [3][4], Celluar Automata (CA) [5], Probabilistic Cellular Automata (PCA) [6] and Hierarchical Automata (HA) [3]. No more details on the definitions of these types of automata are given here due to the limitation of space.
\newtheorem{theorem}{Definition}[section]
\begin{theorem}[Mimic Automata, (MA)]\label{MA}
	A mimic automaton $ A $ is a six-tuple $ (A_{HA},\gamma,A_{SA},C,A_{CA},SA\leftrightarrow CA) $, where
	\begin{itemize}
		\item $ A_{HA} $ is a set of $ HA $;
		\item $ \gamma $ is a set of Composition Functions;
		\item $ A_{SA} $ is a set of $ SA $;
		\item $ C $ is a set of configurations of $ HA $;
		\item $ A_{CA} $ is a set of $ CA $;
		\item $ SA\leftrightarrow CA $ determines the following two ways of combination between $ SA $ and $ CA $. And these two ways can recursively call each other.
	\end{itemize}

1) use $ SA $ to express $ CA(SA\leftarrow CA) $, we have $ \Phi_{onestep}\stackrel{r}{=}final(\delta) $, where $ \Phi_{onestep} $ means one step of the running of $ CA $, $ final(\delta) $ means one running of $ SA $, and $ \stackrel{r}{=} $ denotes that the step of the running of $ CA $ consume the same time with the running of $ SA $.
	
2)	use $ CA $ to express $ SA(SA\rightarrow CA) $,we have $ \delta_{onestep}\stackrel{r}{=}final(\Phi) $, where $ \delta_{onestep} $ means one step of the running of $ SA $, $ final(\Phi) $ means one running of $ CA $, and $ \stackrel{r}{=} $ denotes that the step of the running of $ SA $ consume the same time with the running of $ CA $.
\end{theorem}

From the definition~\ref{MA}, a mimic automaton is composed of sequential automata, cellular automata and hierarchical automata according to some certain logical relations, where sequential automata express state transitions, cellular automata express dynamic heterogeneity (dynamic reconstruction of execution body), hierarchical automata express granularity (different granularity of execution bodies and different levels of state transitions). 

If a cellular automaton is replaced by a random cellular automaton, it can describe the stochastic dynamic reconstruction of execution body. Therefore, a mimic automaton mathematically describes the state transitions in dynamic, heterogeneous and random architectures at different granularities. In other words, the combination of sequential automata, cellular automata, random cellular automata and hierarchical automata makes a mimic automaton show the characteristics of the mimic variants. It should be note that the core properties and characteristics of mimic computing are dynamic, heterogeneous, random and the mimic variants.
\section{An application: mimic automata model of DHR structures}
A mimic automaton is employed to construct the formal model for Dynamical Heterogeneous Redundant (DHR) structure in this section. And DHR structure is the core architecture of network space mimic defense [7].

A DHR structure is illustrated in the left subfigure of Figure 1. The DHR structure can be modeled using the one-dimensional cellular automata in a mimic automaton, as shown in the right subfigure of Figure 1. In this figure, the scheduling algorithm schedules a number of execution bodies in the way of dynamical reconstruction, which perform some computational tasks. The relationship between the DHR structure and a cellular automata is summarized as follows: a cell unit of the automaton represent an execution bodies, the transition function $ \Phi $ of the automaton represent the scheduling algorithm, See Figure 1 for more details.

By using cellular automata to express the dynamic DHR structures, one can use a mimic automaton to formally describe the computation of the execution body in the following way: (1) The current state of a cell in a cellular automaton corresponds to a sequential automaton in a hierarchy of a hierarchical automaton; (2) the input of the DHR structure corresponds to the input of the sequential automaton; (3) the output of the DHR structure corresponds to the output of the sequential automaton. It should be noted that, the state of the cellular automaton remains constant throughout the run of the sequential automaton.
\begin{figure}
\centering
\scalebox{0.4}{\includegraphics{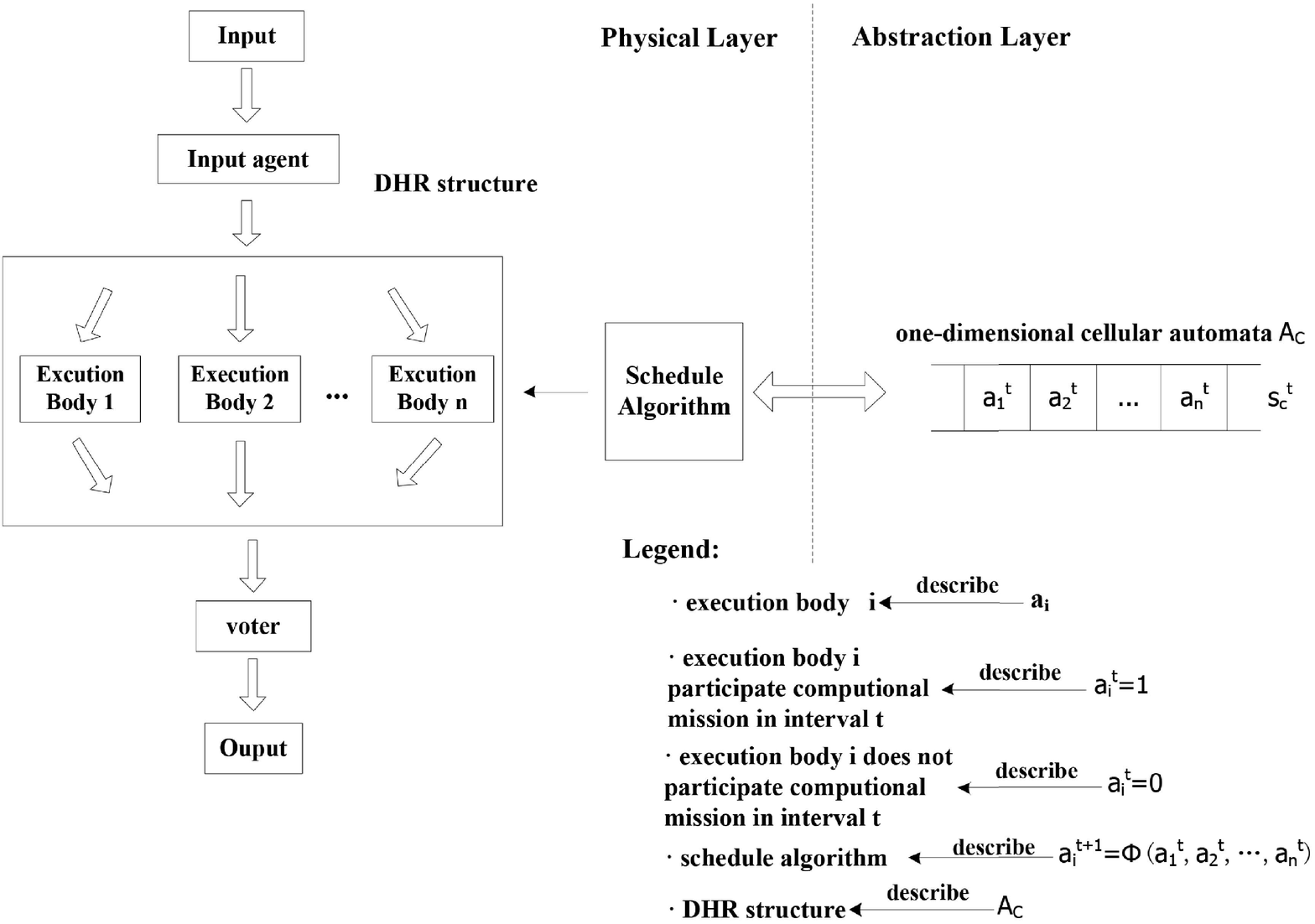}}
\caption{An one-dimensional cellular automaton for modeling a DHR structure}
\end{figure}

In a more complex DHR structure, different DHRs may be organized in a logical relationship. The left part of Figure 2 shows an example. Compared with the left part of Figure 1, the left part of Figure 2 is composed of the two serial DHR structures. The C{\scriptsize 1} sub-tree and C{\scriptsize 2} sub-tree in the right part of Figure 2 establish the formal model for the above two DHRs, respectively. And the components in the first three levels from top to down of the right part of Figure 2 describe the serially sequential relationship between the two DHRs. The SA and the CA are nested with each other in two layers, as shown in the right part of Figure 2. In the inside nest, one-time runing of the SA corresponds to a state transition of the CA. In contrast, one-time running of the CA corresponds to a state transition of the SA in the outside nest. In this example, the HA describes the different granularities, the CA describes the dynamic reconstruction of the execution bodies at a given granularity, and the SA describes the state transitions under the circumstance of a given granularity and a given execution body. A mimic automaton which is composed of the above three types of automata establishes a formal model for a complex DHR structure in this way.
\begin{figure}
	\raggedleft
	\scalebox{0.4}{\includegraphics{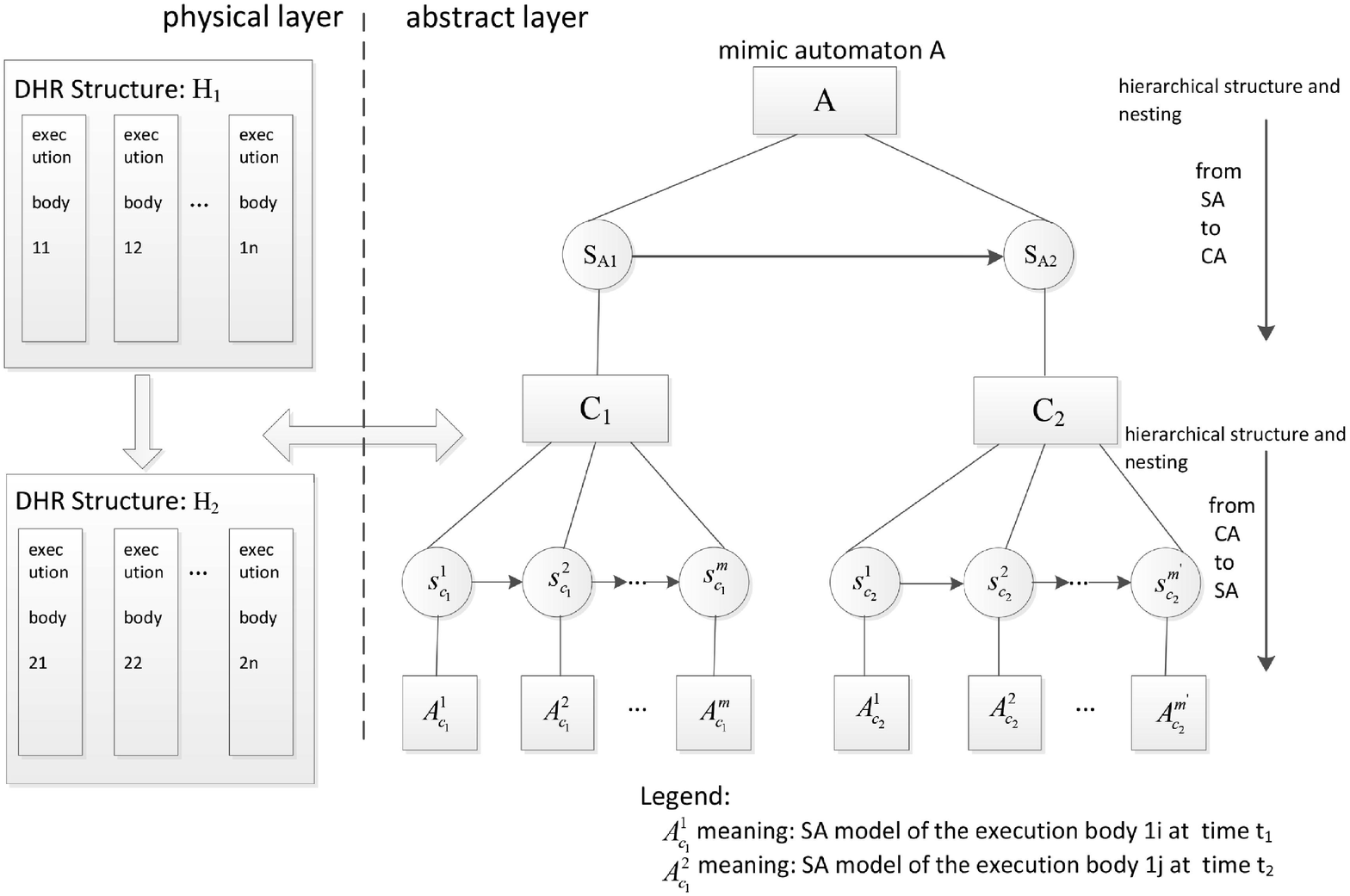}}
	\caption{A mimic automaton for modeling a complex DHR structure}
\end{figure}
\section{The MA model checking}
There are many studies which deal with SA model checking, CA model checking, HA model checking or PA model checking, such as [8], [9], [10], [11] and [12]. Therefore, the general idea about the MA model checking is gotten, by dividing it into the above four kinds of model checking, as shown in Figure 3. No more details are given here due to the limitation of space. 
\begin{figure}
	\centering
	\scalebox{0.39}{\includegraphics{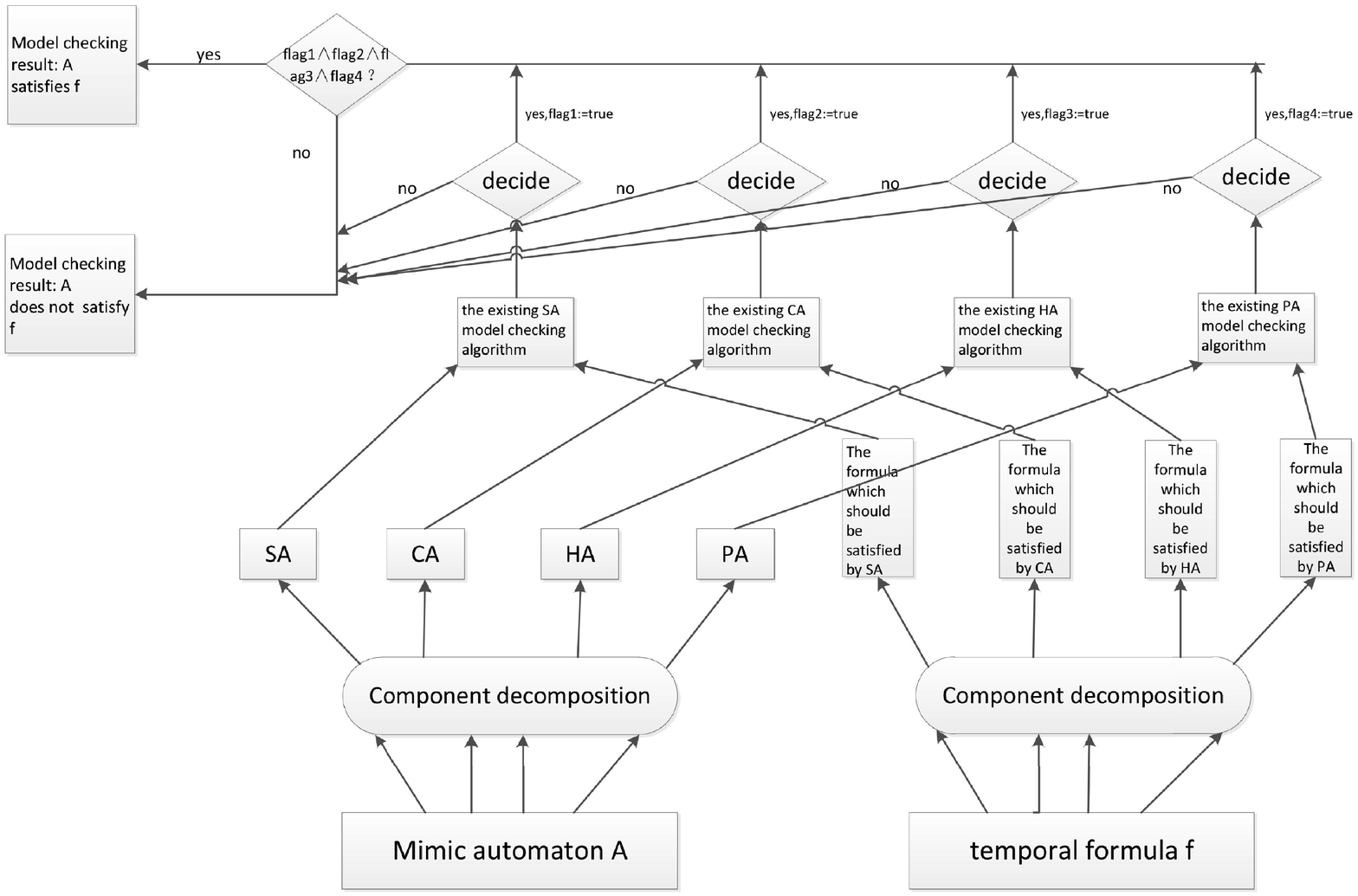}}
	\caption{A flow diagram of MA model checking}
\end{figure}
\section{The malware detection based on MA model checking}
In classical computing, the model checking technique has been employed to detect the malwares [13]. In comparison, the mimic model checking technique can be also applied to Mimic Malware Detection (MMD) on some mimic computers. The principle of a MMD algorithm is depicted in Figure 4.
\begin{figure}
	\raggedleft
	\scalebox{0.55}{\includegraphics{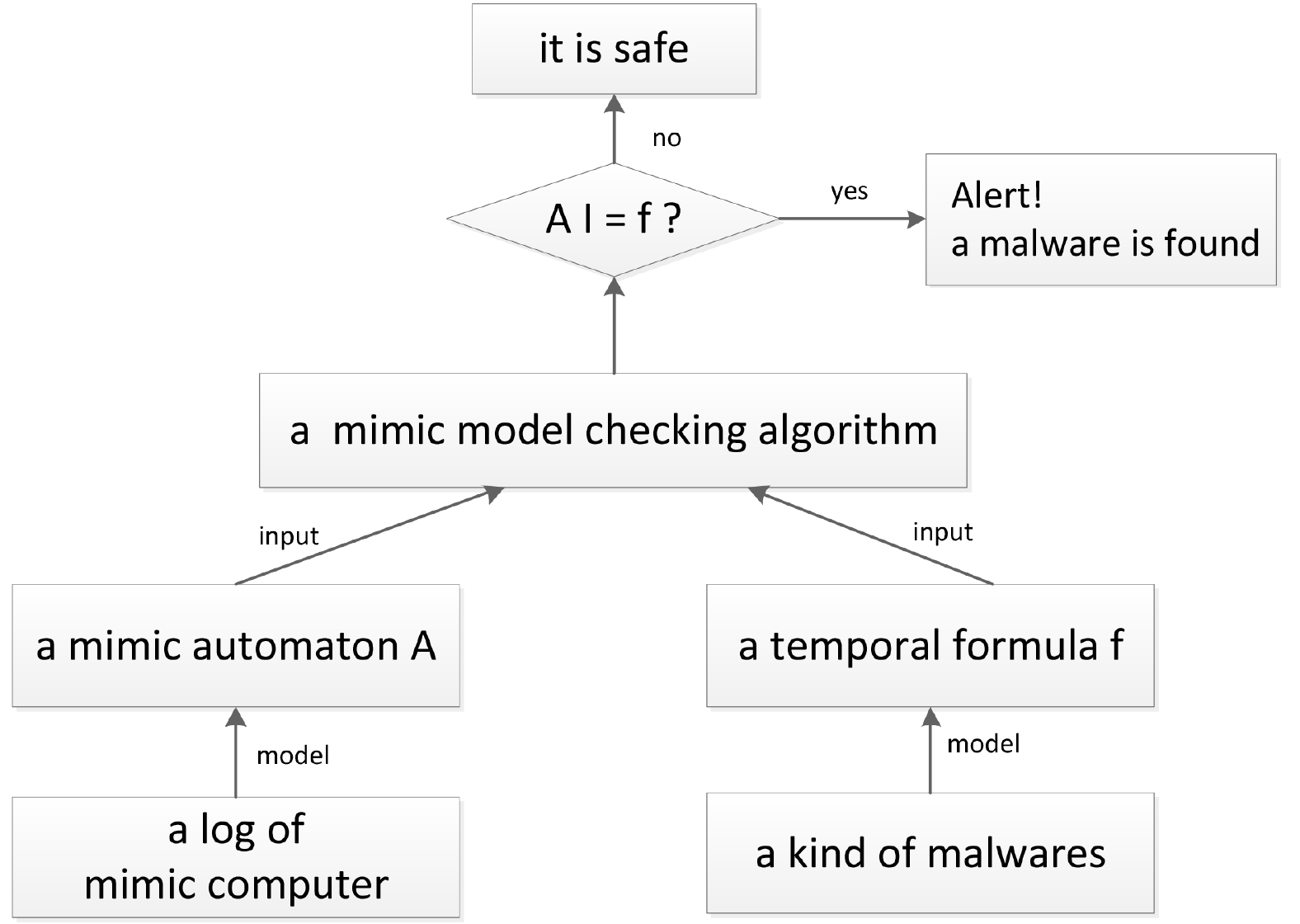}}
	\caption{The principle of a MMD algorithm}
\end{figure}
\section{Conclusions}
The new automaton, i.e., mimic automaton, can be employed to establish a preliminary formal model for mimic computing. With this tool at hand, one can artificially, semi-artificially or  automatically conduct formal analysis for mimic computing systems. This is a prospect of the application of this work.
\section*{Acknowledgements}
This work has been supported by the Natural Science Foundation of China under Grant No.U1204608 as well as China Postdoctoral Science Foundation under Grant No.2012M511588 and No.2015M572120.


\begin{thebibliography}{8}
	\bibitem{JiangxingWU1} Jiangxing WU, Meaning Mimic Computing Mimic Security Defense[J], Telecommunications Science, 2014, 30(7):1-7. (in Chinese)
	\bibitem{JiangxingWU2} Jiangxing WU, Mimic Computing \& Mimic Security [EB/OL], http:$ // $wenku.baidu.com$ /$link?url=wNtBeJMmJoyOoGkW\_{}kCRHFII7w\_{}O
	GNNFCXcp2rpQplaSgqUJUdAjlPGvU8YjZ4t4G-u483lkxIvbq1VLMpfqtAt
	bJwlxB8COz9yxMKXoR3O. (in Chinese)
	\bibitem{TaoYANG} Tao YANG, Tian-yuan XIAO, Lin-Xuan ZHANG, Behavior Modeling for Application Software Using Hierarchical Automata [J], Journal of System Simulation, 2005, 17(4):778-781. (in Chinese)
	\bibitem{HelkeS} Helke S, Kammüller F. Representing Hierarchical Automata in Interactive Theorem Provers[M]// Theorem Proving in Higher Order Logics. Springer Berlin Heidelberg, 2001:233-248.
	\bibitem{WolframS1} Wolfram S. Computation theory of cellular automata[J]. Communications in Mathematical Physics, 1984, 96(1):15-57.
	\bibitem{WolframS2} Wolfram S. Statistical mechanics of cellular automata[J]. Reviews of Modern Physics, 1983, 55(3):601-644.
	\bibitem{HongchaoHU} Hongchao HU, Fucai CHEN, Zhenpeng WANG, Performance Evaluations on DHR for Cyberspace Mimic Defense [J], Journal of Cyber Security, 2016, 1(4)：40-51. (in Chinese)
	\bibitem{JWZ01} J Wang, W Dong, ZC Qi. Slicing hierarchical automata for model checking UML statecharts. Springer Berlin Heidelberg, 2001, 2495: 435-446.
	\bibitem{LLZ09} Li Jing, Li Jinhua, Zhang Fangning. Model checking UML activity diagrams with SPIN. International Conference on Computational Intelligence and Software Engineering, IEEE press, 2009:1-4.
	\bibitem{KD15} Kini, Dileep; Viswanathan, Mahesh, Probabilistic Büchi Automata for LTLGU, Technical Report, http://hdl.handle.net/2142/72686, 2015.
	\bibitem{ZHZ13} ZhouYu , Huang Zhiqiu. Verification of real-time models with hierarchical extensions. ICIC Express Letters. 2013, 7(2):559-564.
	\bibitem{LW11} Gongzheng Lu, Yuejuan Wang. Research on model checking Cellular Automata.
	International Journal of Digital Content Technology and Its Applications. 2011, 5(11):44-51
	\bibitem{Song2012Efficient}
	F. Song and T. Touili, "Efficient malware detection using model-checking," in Proc. 18th
	International Symposium on Formal Methods. Paris, France: Springer,
	2012, pp. 418–433.
		
		
	
		
\end{thebibliography}
\end{document}